\documentclass{jfm}
\usepackage{graphicx}
\usepackage{epstopdf, epsfig}
\usepackage{array}
\usepackage{makecell}
\usepackage{amsmath}
\usepackage{lscape}

\usepackage{rotating}
\usepackage{tabularx}
\usepackage{longtable}
\usepackage{graphicx}
\usepackage{xcolor}
\usepackage{booktabs}
\usepackage{array, multirow, boldline}

\usepackage{amsmath,bm}

\usepackage{subcaption}
\usepackage[labelformat=empty, position=top]{subcaption}
\usepackage[export]{adjustbox}

\usepackage[hidelinks,colorlinks=true,linkcolor=blue,citecolor=blue]{hyperref}
\usepackage{natbib}
\usepackage{apalike}
\usepackage{xpatch}

\usepackage{hyperref}
\hypersetup{
    colorlinks=true,
    citecolor=blue
}

\makeatletter
\xpatchcmd\NAT@citex
 {%
  \@citea\NAT@hyper@{%
    \NAT@nmfmt{\NAT@nm}%
    \hyper@natlinkbreak{\NAT@aysep\NAT@spacechar}{\@citeb\@extra@b@citeb}%
    \NAT@date
  }%
 }
 {%
  \@citea
  \NAT@nmfmt{\NAT@nm}%
  \NAT@aysep\NAT@spacechar
  \NAT@hyper@{\NAT@date}%
 }
 {}{}
\xpatchcmd\NAT@citex
 {%
  \@citea\NAT@hyper@{%
    \NAT@nmfmt{\NAT@nm}%
    \hyper@natlinkbreak{\NAT@spacechar\NAT@@open\if*#1*\else#1\NAT@spacechar\fi}%
    {\@citeb\@extra@b@citeb}%
    \NAT@date
  }%
 }
 {
  \@citea
    \NAT@nmfmt{\NAT@nm}%
    \NAT@spacechar\NAT@@open\if*#1*\else#1\NAT@spacechar\fi
    \NAT@hyper@{\NAT@date}%
 }
 {}{}
\makeatother

\newcommand{\md}{\mathrm{d}}

\shorttitle{Experimental observation for a confined bubble moving in shear-thinning fluids}
\shortauthor{S. Chun, B. Ji, Z. Yang, V.K. Malik and J. Feng}

\title{Experimental observation of a confined bubble moving in shear-thinning fluids}

\author{SungGyu Chun\aff{1},
  Bingqiang Ji\aff{1},
  Zhengyu Yang\aff{1},
  Vinit Kumar Malik\aff{1},
 Jie Feng\aff{1,2},
 \corresp{\email{jiefeng@illinois.edu}}}

\affiliation{\aff{1}Department of Mechanical Engineering and Science, University of Illinois at Urbana-Champaign, Urbana, IL 61801, USA
\aff{2}Materials Research Laboratory, University of Illinois at Urbana-Champaign, Urbana, IL 61801, USA}

\begin{document}

\maketitle

\begin{abstract}
The motion of a long gas bubble in a confined capillary tube is ubiquitous in a wide range of engineering and biological applications. While the understanding of the deposited thin viscous film near the tube wall in Newtonian fluids is well developed, the deposition dynamics in commonly encountered non-Newtonian fluids remains much less studied. Here, we investigate the dynamics of a confined bubble moving in shear-thinning fluids with systematic experiments, varying the zero-shear-rate capillary number $Ca_0$ in the range of $O(10^{-3}-10^2)$ considering the zero-shear-rate viscosity. The thickness of the deposited liquid film, the bubble speed and the bubble front/rear menisci are measured, which are further rationalized with the recent theoretical studies based on appropriate rheological models. Compared with Newtonian fluids, the film thickness decreases for both the carboxymethyl cellulose and Carbopol solutions when the shear-thinning effect dominates. We show that the film thickness follows the scaling law from \citet{aussillous2000quick} with an effective capillary number $Ca_e$, considering the characteristic shear rate in the film as proposed by \citet{picchi2021motion}. $Ca_e$ is calculated by the Carreau number and the power-law index from the Carreau-Yasuda rheological model. The shear-thinning effect also influences the bubble speed and delays the transition to the parabolic region in the bubble front and rear menisci. In particular, a high degree of undulations on the bubble surface results in intricate rear viscosity distribution for the rear meniscus and the deviation between the experiments and theory may require a further investigation to resolve the axial velocity field. Our study may advance the fundamental understandings and engineering guidelines for coating processes involving thin-film flows and non-Newtonian fluids.

\end{abstract}

\begin{keywords}

\end{keywords}

\section{Introduction}

The transport of long gas bubbles or liquid drops in confined geometries plays an important role in many engineering and biological settings, such as enhanced oil recovery \citep{tran2016single, grassia2019motion, majeed2021review}, coating processes \citep{yu2017armoring, jeong2020deposition}, drug delivery \citep{hernot2008microbubbles,gao2016controlled}, biomechanics and biomedical devices \citep{clanet2004motion, chao2020evolution, ma2020dimension, li2021flow}. When such a long bubble of length $L >> R$ translates at a constant speed $U$ in a circular capillary of radius $R$, the bubble forms a symmetrical bullet shape, commonly called a Taylor bubble, and a thin film of liquid is generated between the bubble and capillary. Quantifying the deposition of this liquid film and its relationship with the bubble speed, fluid properties, and channel geometries provides crucial information for determining the mass, momentum, and heat transport in a wide range of multi-phase flow scenarios, and thus remains as a research focus for decades.

The deposition of Newtonian fluids has been extensively studied regarding various geometries and fluid properties. Pioneering investigations on this topic were conducted by \citet{bretherton1961motion} and \citet{taylor1961deposition}. For a long bubble translating in confined geometries with small dimensions where gravity plays a negligible role,  the dynamics are characterized by the interplay between the viscosity and surface tension, as captured by the definition of the capillary number, $Ca = {\mu}U/\sigma$, where $\mu$ is the fluid viscosity and $\sigma$ is the surface tension \citep{aussillous2000quick,jeong2020deposition}. For $Ca \ll 1$, \citet{bretherton1961motion} found that the thickness of the thin liquid film $h$ scales as $h/R \sim Ca^{2/3}$, in regimes where inertia effects are negligible compared to surface tension and viscous effects. This relation was later extended to cover the range of $Ca < 2$ in the scaling analysis of \citet{aussillous2000quick}, where the radius of curvature of the static meniscus was accounted as $R - h$ rather than $R$, yielding a semi-empirical equation as : 
        \begin{equation}
           \dfrac{h}{R}=\dfrac{1.34Ca^{2/3}}{1+2.5\times1.34Ca^{2/3}}.
               \label{AQ}
        \end{equation}

So far, the fluid deposition by a confined bubble in non-Newtonian fluids has been much less understood, although a lot of working fluids, such as polymer solutions, colloidal suspensions, and biologically relevant fluids, show non-Newtonian behaviors in many practical applications \citep{moreira2020isolated, li2021flow,abishek2015dynamics,zhao2021hydrodynamics}. Resolving the hydrodynamics of a confined bubble in non-Newtonian fluids presents more complexities than in Newtonian cases because of the spatial and temporal changes of shear stress and the corresponding variations of the rheological properties. Table \ref{tab:kd} lists a summary of the experimental, numerical, and theoretical studies about the liquid film deposition dynamics in non-Newtonian fluids. In particular, many prior studies assumed a simple power-law model with a stress/shear rate relationship as $\tau={\kappa}\dot{\gamma}^{n_p}$ (where $n_p$ is the power-law index, $\dot{\gamma}$ the shear rate, and $\kappa$ the consistency factor in Pa s$^{n_p}$) to represent the shear-thinning/thickening fluids \citep{kamicsli2001gas, de2007numerical}, and suggested that the liquid film thickness scales as \citep{gutfinger1965films, hewson2009model}:
   \begin{equation}
             \dfrac{h}{R} \sim{} \hat{Ca}^{2/(2n_p+1)} \quad \text{with}\quad \hat{Ca} = \kappa(U/R)^{n_p}/(\sigma/R),
               \label{modieqn}
        \end{equation}where $\hat{Ca}$ is a modified capillary number. Nevertheless, the power-law model fails to reproduce the low-shear-rate viscosity plateau and has a well-known singularity at zero-shear-rate, leading to an inaccurate velocity profile in multiple free-surface flow scenarios \citep{bird1987dynamics,myers2005application, hewson2009model, picchi2017characteristics}. 

\begin{table}
\renewcommand{\arraystretch}{3}
  \begin{center}
\def~{\hphantom{0}}
\begin{adjustbox}{max width=\linewidth}
  {\begin{tabular}{lcccc}
  
\textbf{Reference}  & \textbf{Range of $Ca_0$}   & \textbf{Focus}  & \textbf{\makecell{Fluid/\\Rheological model/\\Channel geometry}}& \textbf{\makecell{Scaling law for film thickness}}  \\[5pt]


    
     \citet{ro1995viscoelastic}  & $Ca_0 \leq 1$ & Theor. & \makecell{Viscoelastic/\\ Oldroyd-B/\\Hele-Shaw cell}&    \makecell{$\dfrac{h}{H}=(1.337-0.08346m^2+O(m^4))Ca_0^{2/3}$\\$-(0.12-0.027k^2+O(k^4))(1-S)WiCa_0^{1/3}$\\$+O(\delta^2Ca_0^{2/3})$}\\

     \citet{gauri1999motion}  & $10^{-2}  \leq Ca_0 \leq 10^{2}$ & Exptl. & \makecell{Boger/\\ 4-mode Giesekus/\\Circular}& (N/A)\\
     
       \citet{kamicsli2001gas}  & $Ca_0 \leq 6$ & Exptl.  & \makecell{Shear-thinning \\and viscoelastic/\\Power-law/\\Circular and rectangular}& (N/A)\\
     
     \citet{yamamoto2004gas}  & $10^{-2} \leq Ca_0 \leq 6$ &  Exptl.    &\makecell{Shear-thinning \\and viscoelastic/ \\Power-law/\\Circular}& (N/A)\\
     
       \citet{de2007numerical}  & $10^{-1} \leq Ca_0 \leq 10$ & Numer.  & \makecell{Shear-thinning\\ and viscoplastic/ \\Power-law/\\Circular} & (N/A)\\
     
      \citet{hewson2009model}  & $10^{-4} \leq Ca_0 \leq 10$ & Theor.  & \makecell{Shear-thinning/ \\Power-law \\and Ellis/\\Circular} & $\dfrac{h}{R} \sim \hat{Ca}^{2/2n_p+1}$\\
      
\citet{boehm2011experimental}   &$10^{-2} \leq Ca_0 \leq 50 $ & Exptl.  & \makecell{Viscoelastic/\\Single-mode Giesekus/\\Square} & (N/A)\\
     \citet{laborie2017yield}   & $Ca_0 \leq 3$   &  Exptl.  & \makecell{Yield stress/\\Herschel-Bulkley/\\Circular}  &  \makecell{$\hat{Ca}_{HB}=a\Big(\dfrac{h}{R}\Big)^{n_{HB}+1/2}\Bigg(\dfrac{2}{1-b\dfrac{h}{R}}-\dfrac{1}{1-\dfrac{h}{R}}\Bigg)^{3/2}$\\$-B\Big(\dfrac{h}{R}\Big)^{n_{HB}}$} \\
     

    \citet{sontti2017cfd}   & $10^{-3} \leq Ca_0 \leq 10^{-2} $ & Numer.  & \makecell{Shear-thinning/\\Power-law/\\Circular} & (N/A)\\
      
    \citet{moreira2020isolated}   & $10^{-2} \leq Ca_0 \leq 0.17$ & Numer.  & \makecell{Shear-thinning \\and viscoelastic/\\Carreau-Yasuda/\\Circular} & (N/A) \\
        
    \citet{zhao2021hydrodynamics} & $10^{-3} \leq Ca_0 \leq 0.006$ &  Exptl.  & \makecell{Viscoelastic/\\Power-law/\\Rectangular} & \makecell{$\dfrac{h}{W} = -0.141+10.9Ca_e^{2/3}$}\\
    
      \citet{picchi2021motion} & $Ca_0 \leq 0.1$ & Theor.  & \makecell{Shear-thinning/\\Ellis/\\Planar two plates} & \makecell{$\dfrac{h}{R} \sim Ca_e^{2/3}$} \\
      
      Present work (2022) & $ O(10^{-3}) \leq Ca_0 \leq O(10^2)$ &  Exptl.  & \makecell{Shear-thinning/\\Carreau-Yasuda/\\Circular} &
      \makecell{$\dfrac{h}{R}=\dfrac{1.34Ca_e^{2/3}}{1+2.5\times1.34Ca_e^{2/3}}$}
  \end{tabular}}
  	\end{adjustbox}
  \caption{Chronological selection of previous experimental, numerical, and theoretical studies on the liquid film thickness of a long bubble translating through non-Newtonian fluids in confined geometries. The non-dimensional groups appearing above are defined as follows: $Ca_0 = {\mu_0}U/\sigma$, $Wi = {\lambda}U/H$, $\hat{Ca}_{HB} = k_{HB}(U/R)^{n_{HB}}/(\sigma/R)$, and $Ca_e = {\mu_e}U/\sigma$. Here, $\mu_0$ is the zero-shear-rate viscosity, $\lambda$ is the relaxation time, $H$ is the half of the gap in the Hele-Shaw cell, $k_{HB}$ and $n_{HB}$ are the consistency factor in Pa s$^{n_{HB}}$ and the power-law index of the Herschel Bulkely model, respectively, and $\mu_e$ is the effective viscosity defined as $\mu_e=\mu(\dot{\gamma}={U}/{h})$. In \citet{ro1995viscoelastic}, $m$ and $k$ represent the degree of shear and normal stress thinning, respectively, $S$ is the ratio of the solvent viscosity to the sum of the polymer and solvent viscosity, and $\delta$ is the ratio between $\lambda$ and the characteristic residence time in the gap. In \citet{laborie2017yield}, $a$ and $b$ are the fitting parameters and $B$ is the non-dimensional number comparing the yield stress to the capillary pressure. In \citet{zhao2021hydrodynamics}, $W$ is the width of the rectangular microchannel.} 

  \label{tab:kd}
  \end{center}
\end{table}
 
Therefore, recent studies have focused on developing more generalized scaling laws to overcome the limitation of the power-law model. Using the Carbopol solution (1.1 wt\%) which shows strong yield-stress effect, \citet{laborie2017yield} experimentally investigated the deposition of the yield-stress fluid in circular channels and developed a semi-empirical scaling law for the film thickness, considering the competition of the yield stress, the capillary pressure, and the viscous stress (see Table \ref{tab:kd}). Additionally, a recent study by \citet{picchi2021motion} investigated the motion of a Taylor bubble through an Ellis fluid and identified a scaling law of the film thickness with the generalized effective viscosity defined by the characteristic shear rate in the liquid film. Several recent studies also showed that the predictions using the power-law model can be erroneous at flow settings with low-shear-rate region and have a small range of applicability compared to the more accurate Ellis and Carreau-Yasuda models \citep{ moukhtari2018semi, picchi2021motion, boyko2021flow}. However, aforementioned studies are still limited to mostly numerical or theoretical perspectives, and a systematic experimental verification of the theoretical predictions is still lacking.

              

In this work, we report an experimental investigation for the effect of shear-thinning rheology on the film deposition dynamics, bubble speed, and bubble shape variations when a bubble is moving in a circular capillary tube filled with non-Newtonian fluids. We consider the range of the zero-shear-rate capillary number, $Ca_0 = {\mu_0}U/\sigma$ where $\mu_0$ is the zero-shear-rate viscosity for non-Newtonian fluids, over six orders of magnitude ($7\times10^{-3} < Ca_0 < 830$). In § \ref{Material and experimental setup}, we provide the experimental framework to measure the film thickness and bubble shape. The generalized Carreau-Yasuda model is used to describe the full range of rheological properties for carboxymethyl cellulose and Carbopol solutions (§ \ref{Rheological properties of CMCell and Carbopol solutions}). Scaling laws for the film thickness based on different rheological models are compared (§ \ref{Scaling of the film thickness with the modified Capillary number} and \ref{Scaling of the film thickness with the effective Capillary number}), showing the film thickness and the bubble speed (§ \ref{Scaling of the bubble speed with the effective Capillary number}) scale with an effective capillary number, $Ca_e$ containing two dimensionless numbers that describe the fluid rheology. Finally, the bubble front and rear menisci are experimentally characterized and further compared with the lubrication theory of \citet{picchi2021motion} (§ \ref{Characteristics of the bubble front and rear meniscus}).

\section{Material and experimental setup}
\label{Material and experimental setup}

We use carboxymethyl cellulose (CMCell, Sigma Aldrich) and Carbopol (981, Lubrizol) solutions with different mass fractions as non-Newtonian fluids, and pure glycerin ($\mu$ = 0.87 Pa~s, $\sigma$ = 63.4 mN/m, Fisher Scientific) as a Newtonian fluid for a baseline comparison. CMCell consists of bentonite as a major component \citep{benchabane2008rheological}, while Carbopol is composed of polyacrylic acid resins \citep{laborie2017yield}. We prepare the CMCell solutions with four different mass fractions (0.5, 1.0, 1.5, and 2.0 wt\%), and the Carbopol solutions with three different mass fractions (0.1, 0.2, and 0.5 wt\%, neutralized using 1 M sodium hydroxide). No elastic behaviors are expected in such a low CMCell mass fraction range given the shear rates in the current experiments \citep{ghannam1997rheological,benchabane2008rheological}. All aqueous solutions were prepared by gradually dissolving a known weight of powders into deionized water in a cylindrical beaker with a continuous stirring. The mixing was maintained for 24 h until clear and homogeneous solutions were produced. The surface tension of the working fluid is determined by the pendant drop method \citep{rotenberg1983determination,song1996determination}.

\begin{figure}
  \centerline{\includegraphics[width=\linewidth]{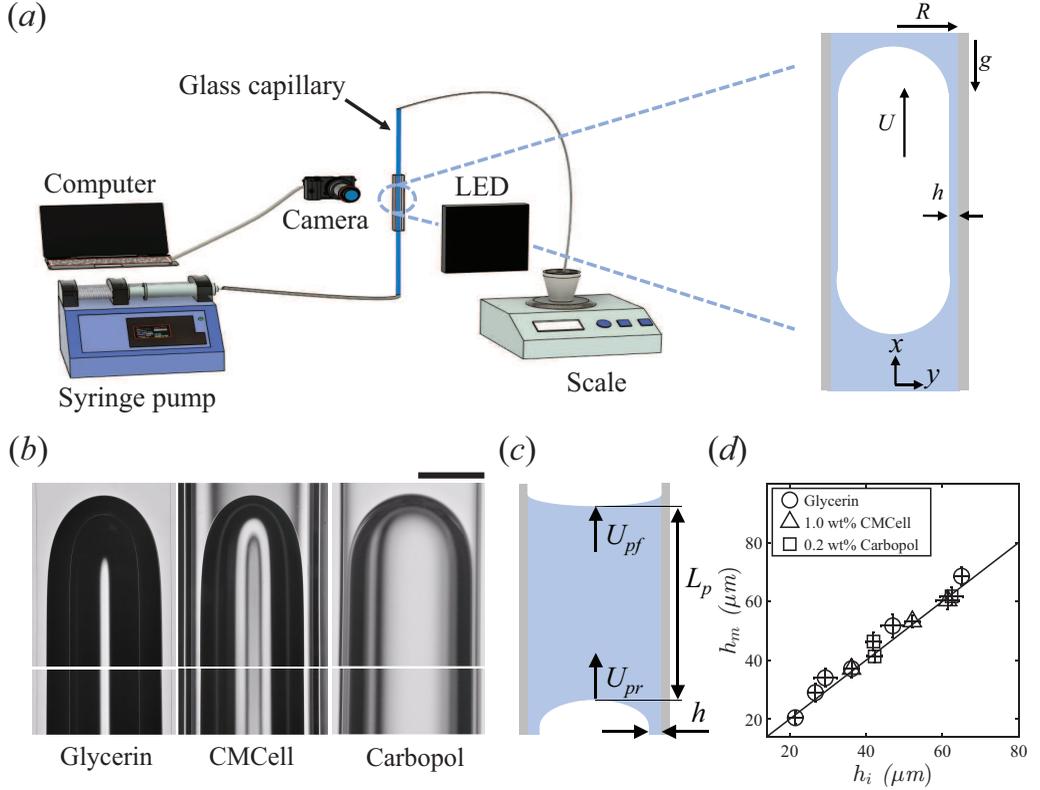}}
  \caption{(\textit{a}) Schematic of the experimental configuration. A cylindrical glass tube (with an inner radius of 0.47 mm) is filled with a sample solution (e.g. glycerin, carboxymethyl cellulose (CMCell; 0.5, 1.0, 1.5, and 2.0 wt$\%$), and Carbopol (0.1, 0.2, and 0.5 wt$\%$)). The central part of the circular glass capillary is submerged in a bath of a sample solution to match the refractive index of glass. Inset: schematic of a translating air bubble confined in a circular tube. (\textit{b}) Typical experimental images of a long bubble as it translates in a circular capillary filled with the glycerin (left) at $Ca_0 = 7.48 \times 10^{-2}$, CMCell (1.0 wt$\%$; center) at $Ca_0 = 3.53 \times 10^{-2}$, and Carbopol (0.2 wt$\%$; right) solutions at $Ca_0 = 3.53 \times 10^{-2}$, respectively. Images contain the front meniscus and the middle part of the bubble, where the film thickness is uniform. The scale bar is 0.5 mm. (\textit{c}) Schematic for the mass balance analysis regarding the deposition of a liquid film in a circular capillary tube. (\textit{d}) Comparison of the liquid film thickness obtained by the image visualization, $h_i$ and the mass balance analysis of the liquid plug, $h_m$ (equation (\ref{filmthickness_mass})) for experimental cases over the range of $8\times10^{-3} < Ca_0 < 8\times10^{2}$. All the data lie along the solid line with a slope of a unity.} 
\label{setting}
\end{figure}

\subsection{Experiments}

Bubble motion experiments were performed in a circular glass capillary tube with a length of 500 mm and an internal radius \textit{R} = 0.47 mm. The glass capillary was held vertical and the central length of the capillary was contained in a clear rectangular box filled with the working fluid, in order to decrease refractive index difference compared to the glass as well as the optical distortion from the curvature of the tube wall \citep{yu2017armoring, zhao2018forced}. 

During the experiments, the inlet of the glass capillary tube was connected to a syringe pump (11 Pico Plus Elite, Harvard Apparatus) using a flexible connecting tube. After the glass capillary was pre-filled with a sample solution, a small volume of air was created in the connecting tube. The flow rate was then set to a very small value $\approx$10 $\mu$L/min in order to steadily transfer the long air bubble (of length \textit{L} $>>$ \textit{R}) from a flexible connecting tube into the inlet of the glass capillary. When the front of the bubble reached the inlet, the syringe pump was set to the targeted flow rate accordingly. The optical images of the region of interest (ROI) were recorded at the rate of 60 frames per second using a digital camera (20.9 Megapixel, D7500, Nikon) equipped with a long working distance objective lens (12× zoom lens system, Navitar); see the schematic in Figure \ref{setting}(\textit{a}). The maximum resolution of the image in our experimental configuration is $\approx$1.3 $\mu$m per pixel, leading to a maximum relative error in the film thickness measurement less than 12\%. The average velocity of the bubble can be evaluated by tracking the gas-liquid interface at the bubble front tip using the images taken with a lower magnification. An analytical balance (ME104E, Mettler Toledo) was installed at the outlet of the glass capillary to confirm the flow rate. In the current experiment, $Bo={\rho}gR^2/\sigma < 0.04$ and $Re = {\rho}UR/\mu \ll 1$ so gravity and inertia effects are negligible \citep{atasi2017effect,magnini2019dynamics}.

   \subsection{Measurement of the liquid film thickness}
   
 The liquid film thickness around the bubble is estimated with the image visualization, in which the film profile around the bubble is obtained from the difference between the position of the bubble surface and the tube wall (see Figure \ref{setting}(\textit{b})). Since the length of the bubble is larger than the size for the field of view in the images, we use a time-strip analysis with ImageJ to ensure an accurate measurement of the uniform liquid film around the bubble. 
 
 To confirm the accuracy from the image visualization method, we also measure the film thickness using the mass balance analysis. In the mass balance analysis, the film thickness is determined based on the change of the length of the liquid plug $L_p$. We ensure the initial length of the liquid plug is about $L_p\approx$ 7 cm, so that $L_p >> R$. The plug advances inside the tube, deceasing $L_p$ (see Figure \ref{setting}(\textit{c})) due to the deposition of the film on the tube wall. The moving positions of the front and rear menisci of the liquid plug are analyzed using ImageJ, which determines the velocity at the front and rear menisci of the liquid plug, $U_{pf}$ and $U_{pr}$, respectively, ranging from 0.2 to 50 mm/s. Assuming a homogeneous deposition of the liquid film near the cylindrical capillary tube wall, a mass balance on the moving plug of length $L_P$ yields the relation as \citep{laborie2017yield}
 \begin{equation}
    \dfrac{h}{R}=1-\sqrt{1+\dfrac{1}{U_{pr}}\dfrac{\md L_p}{\md t}}=1-\sqrt{\dfrac{U_{pf}}{U_{pr}}}.
    \label{filmthickness_mass}
\end{equation}

Figure \ref{setting}(\textit{d}) compares the liquid film thickness measured by the image visualization, $h_i$ and the mass balance analysis, $h_m$ when a bubble is translating in the glycerin, 1.0 wt$\%$ CMCell, and 0.2 wt$\%$ Carbopol solutions. The measurement is conducted over a wide range of $Ca_0$ ($8\times10^{-3} < Ca_0 < 8\times10^{2}$). The results show that all the experimental measurement data lie on the line with a slope of a unity, confirming the two methods give the same result on each experiment. We note that the deposited film thickness is observed to be invariant by rotation along the axis of the glass capillary since the shear-thinning effect dominates for the Carbopol solution with low mass fractions (Figure \ref{setting}(\textit{b})), while the experiments performed by \citet{laborie2017yield} showed that the annular Carbopol solution film thickness was non-uniform azimuthally when the yield stress is important. In the followings, we will use the image visualization to determine the film thickness, $h$.

\section{Results and discussion}

 \subsection{Rheological properties of CMCell and Carbopol solutions}
 \label{Rheological properties of CMCell and Carbopol solutions}
\begin{figure} 
        \includegraphics[width=\linewidth]{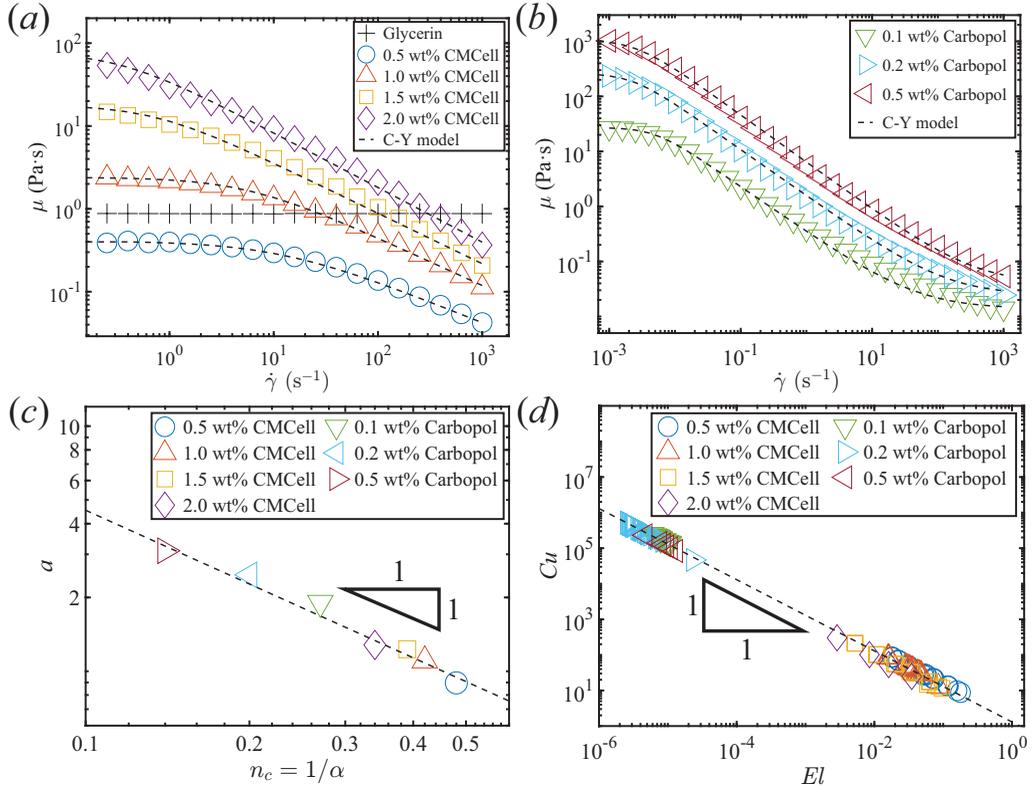}
   \caption{(\textit{a}) Rheogram of the glycerin and carboxymethyl cellulose (CMCell) solutions with different mass fractions: viscosity $\mu$ versus shear rate $\dot{\gamma}$. (\textit{b}) Rheogram of the Carbopol solutions with different mass fractions: $\mu$ versus $\dot{\gamma}$. The dashed lines represent a fitting with the Carreau-Yasuda (C-Y) model. (\textit{c}) Dimensionless parameter $a$ versus the power-law index $n_c$ in the C-Y model or the degree of shear-thinning $\alpha$ in the Ellis model. (\textit{d}) Ellis number $El$ versus Carreau number $Cu$ in the current experiments. Error bars are smaller than the symbols.}
    \label{rhe}
\end{figure}

Rheological measurements of the CMCell and Carbopol solutions are performed with a controlled stress rheometer (DHR-3, TA Instrument) using a parallel-plate geometry (with a diameter of 25 mm) at controlled temperature of 25 $^{\circ}$C. Under a simple shear, the rheological properties of shear-thinning fluids are classically modelled by the power-law model $\mu=\kappa\dot{\gamma}^{n_{p}-1}$ \citep{kamicsli2001gas,de2007numerical}. However, the power-law model cannot predict the viscosity at the low-shear-rate region \citep{picchi2017characteristics,picchi2021motion}, where the viscosity approaches to a constant value, known as the zero-shear-rate viscosity. Instead, the Ellis model \citep{reiner1960deformation} was proposed to capture such a viscosity plateau  with a constitutive equation as
 \begin{equation}
             \dfrac{\mu}{\mu_0}=\dfrac{1}{1+(\tau/\tau_{1/2})^{\alpha-1}},
             \label{Ell}
        \end{equation}where $\tau_{1/2}$ is the shear stress at which the viscosity is half of the Newtonian limit, while $\mu_0$ and $\alpha$ are the zero-shear-rate viscosity and the degree of shear-thinning, respectively. 
        
        Here, we consider the Carreau-Yasuda (C-Y) model, which has been used to describe emulsions, protein solutions, and polymer melts \citep{myers2005application,picchi2017characteristics}. The C-Y model is more convenient for experimental analysis since it expresses the viscosity as an explicit function of the shear rate \citep{carreau1972rheological, morozov2015introduction,pipe2008high}. The constitutive equation of the C-Y model is
       \begin{equation}
             \mu=(\mu_0-\mu_\infty)(1+(\lambda\dot{\gamma})^a)^{(n_c-1)/a}+\mu_\infty,
             \label{Car}
        \end{equation} 
where $\mu_0$, $\mu_\infty$, $n_c$, and $a$ are the zero-shear-rate viscosity, infinite-shear-rate viscosity, the power-law index, and dimensionless parameter, respectively. $\lambda$ is the inverse of a characteristic shear rate at which shear-thinning becomes apparent. Figures \ref{rhe}(\textit{a}) and (\textit{b}) demonstrate that the C-Y model can well capture the rheological behaviors for both the CMCell and Carbopol solutions in the range of shear rates over five orders of magnitude. We note that $\mu_\infty/\mu_0<O(10^{-2})$ and therefore $\mu_\infty$ is neglected in the following discussion. 

A very recent theoretical work by \cite{picchi2021motion} derived the film thickness as a function of the Ellis number $El$ and the degree of shear-thinning $\alpha$ for an Ellis fluid, where $El$ is the ratio between the characteristic shear rate of the fluid and the characteristic shear rate in the liquid film as
\begin {equation}
        El=\dfrac{\tau_{1/2}h}{U\mu_0}.
        \end{equation}In order to compare our experimental measurements with the lubrication theory of \cite{picchi2021motion}, we rewrite the C-Y model with a similar form to the Ellis model as
       \begin{equation}
             \frac{\mu}{\mu_0}=\left(1+{\left({Cu\tilde{\dot{\gamma}}}\right)}^a\right)^{(n_c-1)/a}  \quad \text{with}  
             \enspace  Cu = \dfrac{\lambda{U}}{h} 
             \enspace \text{and} 
             \enspace \tilde{\dot{\gamma}}={\frac{\dot{\gamma}}{U/h}},
             \label{Carreanumber1}
        \end{equation} where the Carreau number $Cu$ is the ratio between the effective shear rate $U/h$ in the film and the crossover strain rate $1/\lambda$ \citep{datt2015squirming} and $\tilde{\dot{\gamma}}$ is the dimensionless shear rate. 
With the rheological data in Figures \ref{rhe}(\textit{a-b}), we used the method of least squares fitting to obtain all the parameters for the Ellis and C-Y models. Notably, the value of $a$ in the C-Y model is chosen to impose $n_c=1/\alpha$ for an analogy between the Ellis and C-Y models, and we find $a\approx0.45 n_c ^{-1}$ as shown in Figure \ref{rhe}(\textit{c}). Next, for all the experimental cases, once the bubble speed and the film thickness are determined, $El$ and $Cu$ are calculated as shown in Figure \ref{rhe}(\textit{d}). In the current work, $Cu$ is found to be inversely proportional to $El$ with an experimentally fitted relation of $Cu\approx1.3El^{-1}$.
The rheological parameters of the CMCell and Carbopol solutions are reported in Table \ref{tab:CMCellcarb}, and we will focus on the effects of $Cu$ and $n_c$ on the deposition dynamics of the working fluids.

We note that Carbopol solutions can exhibit both yield stress and shear-thinning behaviors. However, we use a low mass fraction of Carbopol to diminish the yield stress effect \citep{spiers1975free, ma2015experimental}, so that only the shear-thinning behavior dominates. We further justify this consideration by fitting our rheological data of Carbopol with the Herschel-Bulkley model $\tau=\tau_y+k_{HB}\dot{\gamma}^{n_{HB}}$, where $\tau_y$ is the yield stress and $k_{HB}$ is the consistency factor. A dimensionless number $B=\tau_y R/\sigma$ is suggested to compare the yield stress to the capillary pressure \citep{deryagin1964film, laborie2017yield}. In our experiments, $B$ is $O (10^{-3})$ while $B$ is $O (1)$ in \citet{laborie2017yield}, thus we neglect the yield stress effect in the following discussion.

\begin{table}
  \begin{center}
\def~{\hphantom{0}}
  {\begin{tabular}{lccccccc}
    &         &    \multicolumn{5}{c}{Carreau-Yasuda model} 
      \\\cmidrule(lr){3-7}
       
   \multirow{3}{*}{Fluids}  & $\sigma$    &   $\lambda$   &      $\mu_0$ &$\mu_\infty$  & $n_c$ & $a$   &  $Cu$
      
      \\
       &   (mN/m) & (s)  &  (Pa~s)  &  ($\times10^{-2}$, Pa~s)  & &     &
   \\
   \\
    CMCell 0.5 wt\% &  73.2 $\pm$ 0.3 & 0.1 & 0.41 $\pm$ 0.04   & 0.1 $\pm$ 0.01 & 0.48 & 0.9  &  10.3 - 86.7
   \\ 
     CMCell 1.0 wt\% & 73.2 $\pm$ 0.4 & 0.2 & 2.4 $\pm$ 0.06  & 0.1 $\pm$ 0.01  &  0.42 & 1.1 &  27.1 - 87.5
   \\ 
    CMCell 1.5 wt\% & 67.4 $\pm$ 0.5 & 1.4 & 18.0 $\pm$ 0.08 & 0.9 $\pm$ 0.03 & 0.39 & 1.2  &  11.6 - 214.3
   \\ 
   CMCell 2.0 wt\% & 62.4 $\pm$ 0.2 & 3.0 & 78.1 $\pm$ 0.12  & 0.5 $\pm$ 0.04 & 0.34 &  1.3 & 14.7 - 301.6
   \\ 
 
   Carbopol 0.1 wt\% & 73.7 $\pm$ 0.4   &   300 &   28.3 $\pm$ 0.9 & 1.1 $\pm$ 0.01 &  0.27   &  1.9   & 1.4 - 2.5$\times10^{5}$
   \\
    Carbopol 0.2 wt\% & 71.0 $\pm$ 0.5    &   530       &   263.7 $\pm$ 3.1   & 2.0  $\pm$ 0.03  &  0.20   &  2.5      & 1.6 - 5.2$\times10^{5}$
   \\
  Carbopol 0.5 wt\% & 60.8 $\pm$ 0.3     &  440       &   1100.0$\pm$ 12.1  & 4.0 $\pm$ 0.03  &  0.14   &  3.1       & 0.5 - 2.3$\times10^{5}$
   \\
  
  \end{tabular}}
  \caption{Rheological properties of the CMCell and Carbopol solutions with different mass fractions computed by the C-Y model.}
  \label{tab:CMCellcarb}
  \end{center}
\end{table}

\subsection{Scaling of the film thickness with the modified capillary number}
\label{Scaling of the film thickness with the modified Capillary number}

\begin{figure}
  \centerline{\includegraphics[width=\linewidth]{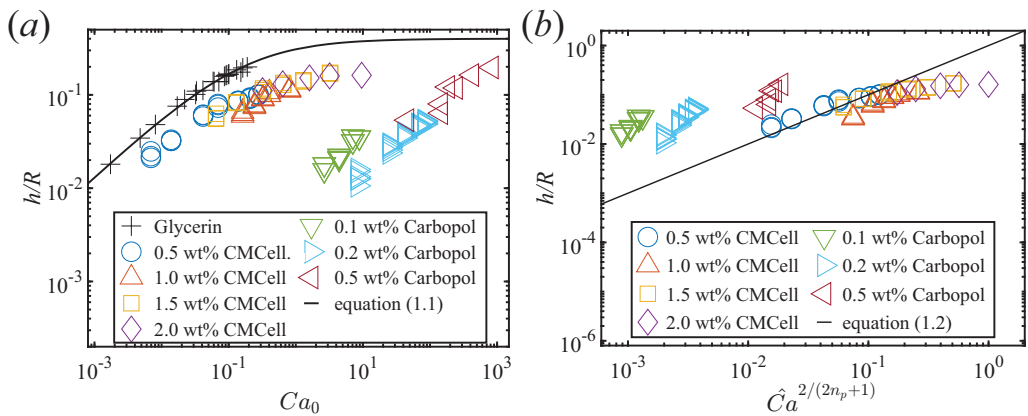}}
  \caption{(\textit{a}) Non-dimensional liquid film thickness \textit{h/R} as a function of $Ca_0$. The black line represents prediction of equation (\ref{AQ}). (\textit{b}) $h/R$ versus $\hat{Ca}^{2/{2n_p+1}}$, where $\hat{Ca}=\kappa(U/R)^{n_p}/(\sigma/R)$, with $\kappa$ and $n_p$ ranging from 0.4 to 5.4 $\textrm{Pa s}^{n_p}$ and 0.16 to 0.51, respectively. The black line represents the prediction of equation (\ref{modieqn}) with a prefactor of 1. The experimental measurements are shown as open symbols, and error bars are smaller than the symbols.}
\label{hrca}
\end{figure}

The film thickness measured in the experiments with the CMCell and Carbopol solutions is shown in Figures \ref{hrca}(\textit{a}) and (\textit{b}), respectively, as a function of the zero-shear-rate capillary number $Ca_0$, as well as the modified capillary number $\hat{Ca}$. As shown in Figure \ref{hrca}(\textit{a}), compared to a Newtonian fluid at the same $Ca_0$, the bubble forms a thinner liquid film in both the CMCell and Carbopol solutions because of the shear-thinning effect. The thinner liquid film formed in the Carbopol solutions compared to that in the CMCell solutions at the same $Ca_0$ results from that the stronger shear-thinning effect in the Carbopol solutions than that in the CMCell solutions, as indicated by lower power-law indices of the Carbopol solutions compared to those of the CMCell solutions in Table \ref{tab:CMCellcarb}.

With $\hat{Ca}$ to compare the shear stress from the power-law model and the capillary pressure, equation (\ref{modieqn}) has been used to predict the film thickness \citep{gutfinger1965films, hewson2009model}. We note that the values of $\kappa$ and $n_p$ for $\hat{Ca}$ are obtained by fitting the rheological data with a power-law model considering the range of shear rates exhibiting a shear-thinning behavior, i.e. $\dot{\gamma}$ = $O(10-10^3$ s$^{-1}$) for CMCell and $O(10^{-1}-10^2$ s$^{-1}$) for Carbopol. As shown in Figure \ref{hrca}(\textit{b}), we report that the data do not precisely follow the power law of $2/(2n_p+1)$ as predicted by equation (\ref{modieqn}). The results imply that all the dynamics of the coating process cannot be captured by using $\hat{Ca}$, since the power-law model alone is not sufficient to describe the rheological behaviors of the working fluids around the bubble. In particular, the deviation of the experimental data for the CMCell solutions is larger than those for the Carbopol solutions when comparing to the prediction of equation (\ref{modieqn}), which can be attributed to the range of the effective shear rates. In the experiments, considering the effective shear rates in the film $\dot{\gamma}=U/h$, we obtain $Cu=O(10-10^2)$ for the CMCell solutions while $Cu=O(10^5)$ for the Carbopol solutions. The experiments for the CMCell solutions include low- to intermediate-shear-rate regions, while the experiments for the Carbopol solutions are performed at intermediate- to comparably high-shear-rate regions. Therefore, the viscosity plateau at low-shear-rate is required to be taken into account to obtain the better prediction for the case of the CMCell solutions, in addition to the power-law dependence at intermediate-shear-rates. Furthermore, in the vicinity of the uniform film thickness region, the fluid is at rest. However, the shear-thinning effect plays an important role at the bubble front meniscus, as the film starts growing rapidly with an increasing shear rate. On the other hand, the local shear rate will further decrease in the region of the re-circulating flow ahead of the bubble. Such a change of the local shear rate at different regions requires an accurate viscosity model for a correct representation of the flow physics, which also highlight the importance of a more realistic rheological model in free surface flow analyses.

\begin{figure}
  \centerline{\includegraphics[width=\linewidth]{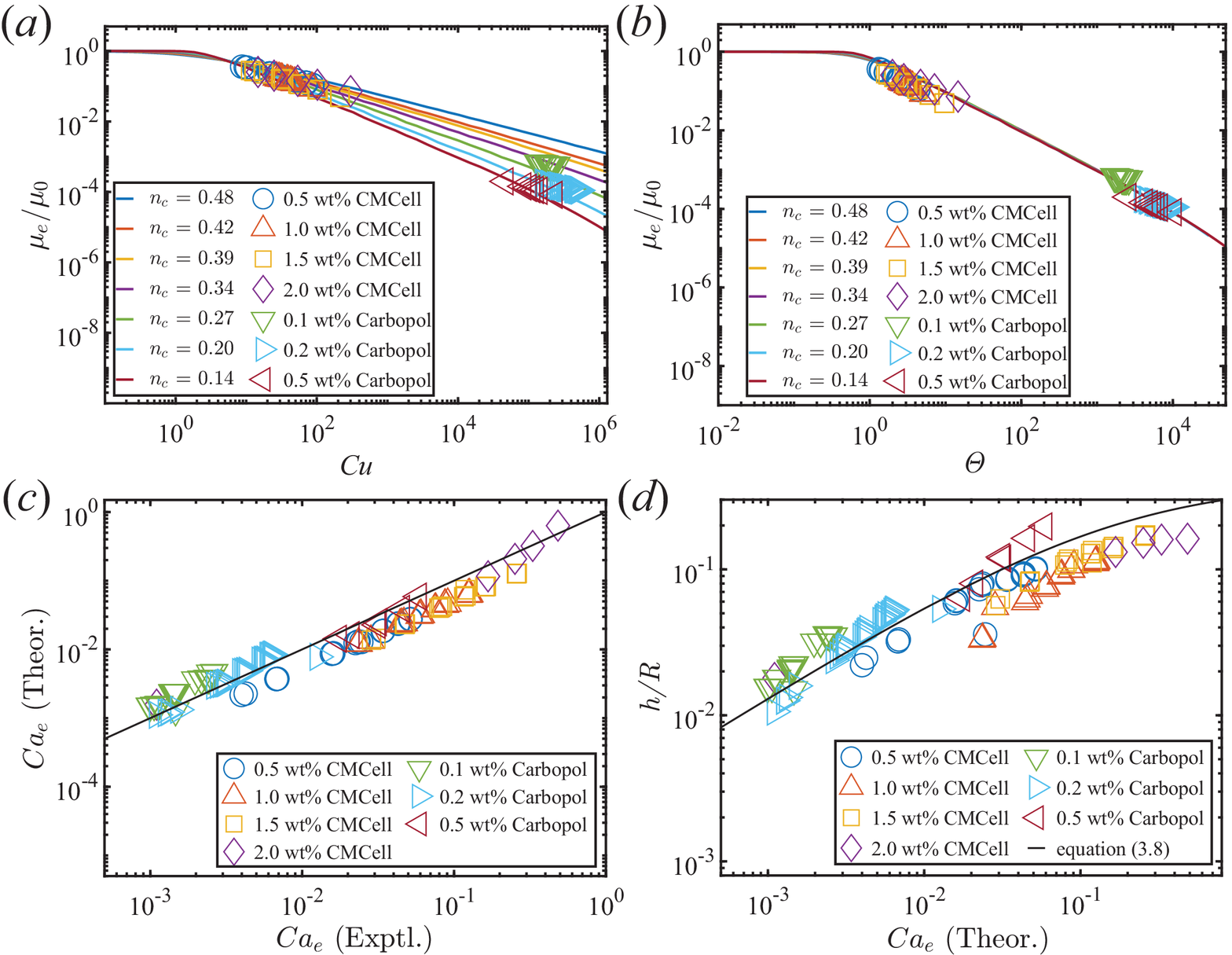}}
  \caption{(\textit{a}) Non-dimensional effective viscosity $\mu_e/\mu_0$ as a function of the Carreau number $Cu$ and the power-law index $n_c$. (\textit{b}) $\mu_e/\mu_0$ of the experimental data as a function of $\Theta$. 
  All the experimental data from the present work show good agreement with the master curve of equation (\ref{master}). (\textit{c}) Comparison of the effective capillary number, $Ca_e$ between the theoretical and experimental results. All the data lie along the solid line with a slope of a unity. (\textit{d}) \textit{h/R} as a function of $Ca_e$. The black line represents prediction of equation (\ref{hRCae}) with $Ca_e$ \citep{aussillous2000quick, picchi2021motion}. Error bars are smaller than the symbols.}
\label{cae}
\end{figure}

\subsection{Scaling of the film thickness with the effective capillary number}
\label{Scaling of the film thickness with the effective Capillary number}

To explore the effect of the shear-thinning rheology on bubble characteristics, the following ordinary differential equation for the bubble profile has been obtained in the theoretical work from \citet{picchi2021motion} considering an Ellis fluid
   \begin{equation}
               \dfrac{\md^3\eta}{\md\xi^3}+\dfrac{3^\alpha}{(\alpha+2){El}^{\alpha-1}}\dfrac{\md^3\eta}{\md\xi^3}\Bigg |\dfrac{\md^3\eta}{\md\xi^3}\Bigg |^{\alpha-1}\eta^{\alpha-1}=\dfrac{\eta-1}{\eta^3},
                \label{ODE}
\end{equation} where $\xi=x/[h(3Ca)^{-1/3}]$ and $\eta=y/h$. Different from the Newtonian case, the bubble profile $\eta$ becomes a function of $\xi$, $\alpha$, and $El$. The two terms in the left-hand side of equation (\ref{ODE}) represent the Newtonian and shear-thinning contributions, respectively. For the uniform film region, the thickness can be obtained by matching the curvature of the parabolic region with the curvature of the bubble spherical cap, $1/R$. 

By introducing the effective capillary number $Ca_e$ that considers both the zero-shear-rate and the shear-thinning effects, we obtain
  \begin{equation}
            \dfrac{h}{R}=P(3Ca_0)^{2/3}=0.643(3Ca_e)^{2/3} \quad \text{with}  \enspace Ca_e=\dfrac{\mu_eU}{\sigma},
                \label{h/R}
\end{equation}
where $\mu_e$ is the effective viscosity defined as $\mu_e=\mu_0(P/0.643)^{3/2}$, and $P$ is the dimensionless curvature related to the capillary pressure as the second derivative of $\eta$ with respect to $\xi$, i.e. $P=\md^2\eta/\md\xi^2$. $P$ is determined by numerically solving equation (\ref{ODE}) using the fully implicit solver $ode15i$ of Matlab and the exact initial conditions given by \citet{picchi2021motion}, when $\eta\gg1$ at the front meniscus. 0.643 is the numerical factor for the Newtonian limit \citep{bretherton1961motion}. In addition, \citet{picchi2021motion} used the numerical results of $P$ from equation (\ref{ODE}) to obtain a master fitting curve of $\mu_e$ as a function of $El$ and $\alpha$. Here, we revise the fitting curve with $Cu$ and $n_c$ considering the experimental rheological data ($Cu\approx1.3{El}^{-1}$ and $n_c=1/\alpha$, § \ref{Rheological properties of CMCell and Carbopol solutions}) as follows:
 \begin{equation}
                \dfrac{\mu_e}{\mu_0}=\left(\dfrac{P}{0.643}\right)^{3/2}=
                \begin{cases}
                    1 & \text{if} \quad Cu \rightarrow 0, \\
                \dfrac{10-7n_c}{4}\left(\dfrac{Cu}{1.3}\right)^{n_c-1} \equiv \Theta & \text{if} \quad Cu >> 1,
              \end{cases}
                \label{master}
        \end{equation} when $Cu \rightarrow 0$, equation (\ref{master}) reduces to a Newtonian case. When $Cu >> 1$, the shear-thinning effect dominates, and thus the viscosity depends on $n_c$ and $Cu$ given that $\mu_\infty$ is neglected in the current experiments. For other practical non-Newtonian fluids with non-negligible $\mu_\infty$, the limit of $Cu \rightarrow \infty$ corresponds to the high-shear-rate viscosity plateau with constant viscosity $\mu_\infty$.


For each experiment, we compute the effective shear rate in the film $U/h$, using the measured bubble speed and film thickness, and thus acquire the effective viscosity $\mu_e/\mu_0$ of the CMCell and Carbopol solutions by using the C-Y model (Figure \ref{rhe}). We find the experimental values of $\mu_e/\mu_0$ agree well with $(P/0.634)^{3/2}$, which is numerically calculated from equation (\ref{ODE}). Indeed, Figure \ref{cae}(\textit{a}) provides a plot of the effective viscosity as a function of $Cu$ and $n_c$ and all the viscosity data collapse around the fitting curve of the master equation (\ref{master}) as shown in Figure \ref{cae}(\textit{b}), which suggests a universal scaling for the effective viscosity to define the effective capillary number. Therefore, the comparison of the effective capillary number $Ca_e$ obtained from equation (\ref{h/R}) against the experimental data show good agreement at the entire range of $Ca_e$ (Figure \ref{cae}(\textit{c})). The smaller values of $\mu_e$/$\mu_0$ for the Carbopol solution demonstrate the higher extent of shear-thinning compared to the CMCell solution, which are also indicated by the higher values of $Cu$ for the Carbopol solutions as shown in Table \ref{tab:CMCellcarb}. We further recast equation (\ref{h/R}) in the following expression proposed by \citet{aussillous2000quick} for the range of $Ca_e$ ($10^{-3}<Ca_e<0.6$) in the current experiments as
  \begin{equation}
               \dfrac{h}{R}=\dfrac{1.34Ca_e^{2/3}}{1+2.5\times1.34Ca_e^{2/3}.
        \label{hRCae}}
        \end{equation}

Figure \ref{cae}(\textit{d}) shows a plot of the non-dimensional liquid film thickness $h/R$ as a function of $Ca_e$. The values of $h/R$ increase consistently with $Ca_e$ for the non-Newtonian fluids, as the viscous effect is increasingly important. The experimental data for both the CMCell and Carbopol solutions agree well with equation (\ref{hRCae}), including the trend in low $Ca_e$ and the saturation behavior at relatively large $Ca_e$. Thus, we demonstrate that the liquid film thickness of the shear-thinning fluids can be estimated with the scaling law proposed by \citet{aussillous2000quick} using $Ca_e$. The better prediction accuracy compared with equation (\ref{modieqn}) also highlights the importance of a practical rheological model. To the best of our knowledge, our work serves as the first experimental validation of equation (\ref{hRCae}) with realistic shear-thinning fluids based on $Ca_e$, which is helpful to assess the true range of applicability for different scaling laws.

\subsection{Scaling of the bubble speed with the effective capillary number}
\label{Scaling of the bubble speed with the effective Capillary number}
                
\begin{figure}
  \centerline{\includegraphics[width=\linewidth]{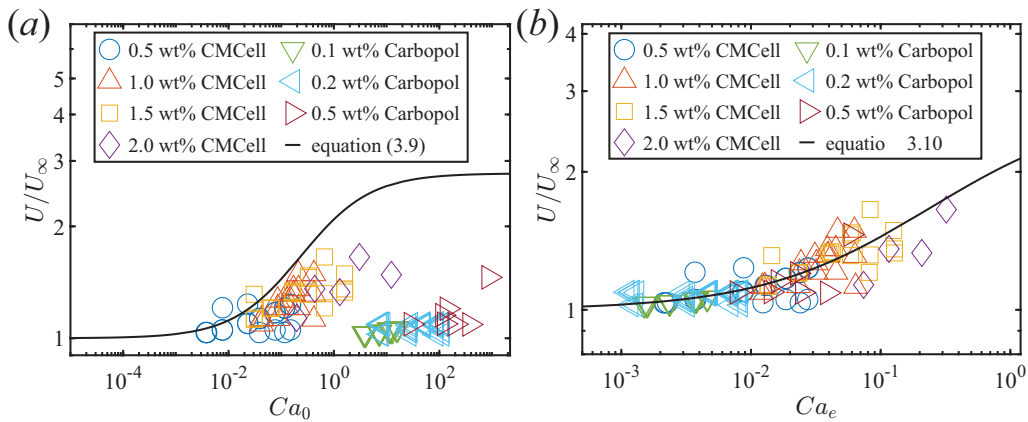}}
  \caption{Ratio of the bubble speed to the average velocity of the fluid far from the bubble $U/U_\infty$, as a function of (\textit{a}) $Ca_0$ and (\textit{b}) $Ca_e$. Error bars are smaller than the symbols.}
\label{ubuf}
\end{figure}
In addition, we experimentally measure the ratio of the bubble speed $U$, to the average velocity of the fluid flowing far from the bubble $U_\infty$. The scaling law for the ratio $U/U_\infty$ can be derived by applying the mass balance near the thin film region in a reference frame moving with the bubble \citep{picchi2021motion}. Using $Ca_0$, we obtain
   {\begin{equation}
               \dfrac{U}{U_\infty}=\dfrac{1}{(1-h/R)^2}=\dfrac{1}{\left(1-\dfrac{1.34Ca_{0}^{2/3}}{1+2.5\times1.34Ca_{0}^{2/3}}\right)^2}.
                \label{ub_uf}
        \end{equation}}
        
        Figure \ref{ubuf}(\textit{a}) shows that the bubble translating 
    in shear-thinning fluids moves slowly compared to a Newtonian bubble with the same $Ca_0$. As $Ca_0$ increases, the deviation between the experimental data and the prediction of equation (\ref{ub_uf}) increases. The slower bubble speed in the Carbopol solution compared to that in the CMCell solution at the same $Ca_0$ results from the stronger shear-thinning effects in Carbopol, as indicated by higher values of $Cu$ and lower values of $n_c$ in Carbopol. With $Ca_e$, equation (\ref{ub_uf}) becomes
            {\begin{equation}
               \dfrac{U}{U_\infty}=\dfrac{1}{\left(1-\dfrac{1.34Ca_{e}^{2/3}}{1+2.5\times1.34Ca_{e}^{2/3}}\right)^2}.
                \label{ube_ufe}
                 \end{equation}}The experimentally obtained $U/U_\infty$ collapse well with equation (\ref{ube_ufe}) (Figure \ref{ubuf}(\textit{b})), showing that $Ca_e$, as a function of $Cu$ and $n_c$, can be used to describe the evolution of the bubble speed with the rheological parameters of the shear-thinning fluids.
        

\subsection{Characteristics of the bubble front and rear menisci}
\label{Characteristics of the bubble front and rear meniscus}

\begin{figure}
  \centerline{\includegraphics[width=\linewidth]{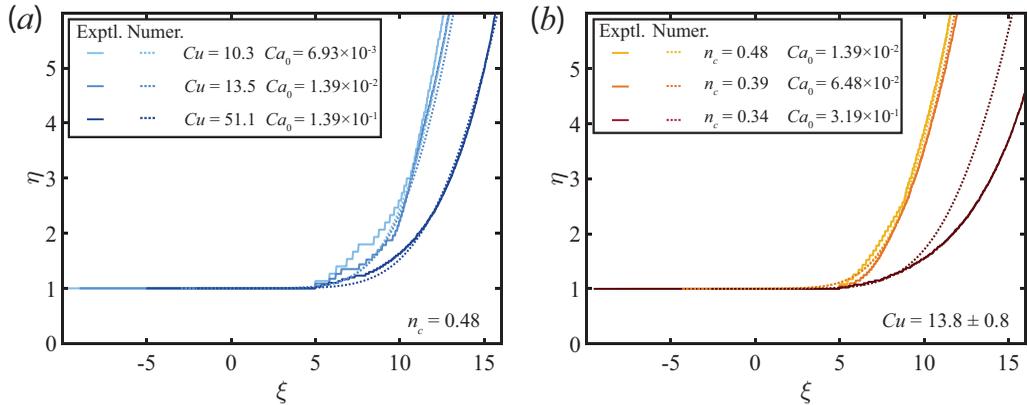}}
  \caption{(\textit{a}) Bubble front meniscus as a function of the Carreau number $Cu$ with the power-law index $n_c = 0.48$. (\textit{b}) Bubble front meniscus as a function of $n_c$ with $Cu = 13.8 \pm 0.8$.}
\label{front}
\end{figure}

\begin{figure}
  \centerline{\includegraphics[width=\linewidth]{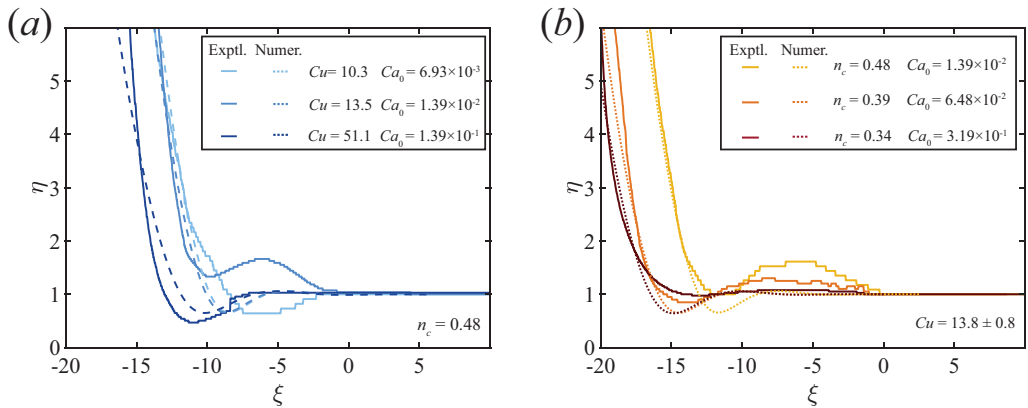}}
  \caption{(\textit{a}) Bubble rear meniscus as a function of the Carreau number $Cu$ with the power-law index $n_c = 0.48$. (\textit{b}) Bubble rear meniscus as a function of $n_c$ with $Cu = 13.8 \pm 0.8$.}
\label{rear}
\end{figure}

We further investigate the shape variations of the bubble translating in the CMCell solution, in which the front and rear menisci of the bubble could be identified clearly compared to those in the Carbopol solution. In this section, the bubble shape profiles near the front and rear menisci are computed by solving equation (\ref{ODE}) to compare with the experimental data. Using the experimentally obtained bubble profiles, we first identify the points, at which the local film thickness increase by one pixel ($\approx 1.3~\mu$m) compared with the uniform film thickness. Such a thickness increase corresponds to $\approx1.7-8.5$\% of the uniform thickness of the deposited film. Then, these points are overlapped with those corresponding to the same thickness increase in the numerically obtained bubble profiles, as shown in Figures \ref{front} and \ref{rear}. We note that the bubble profiles at the front and the rear menisci are solved separately by integrating equation (\ref{ODE}) with a different set of boundary conditions. At the bubble front, we assume that the thin film region extends to $\xi \rightarrow -\infty$ (i.e. $\eta(-\infty)=1$), and the front meniscus is obtained by integrating equation (\ref{ODE}) towards positive $\xi$. However, at the rear meniscus, the thin film region is at $\xi \rightarrow +\infty$, and the profile at the rear meniscus is obtained by integrating equation (\ref{ODE}) towards negative $\xi$ starting from the boundary condition $\eta(+\infty)=1$ \citep{bretherton1961motion, picchi2021motion}.

 For the bubble front meniscus, Figure \ref{front}(\textit{a}) shows good agreement between the experimental results and the numerical predictions \citep{picchi2021motion} of the shape changes when $Cu$ increases at fixed $n_c=0.48$. Although the bubble maintains the rounded shape similar to that in the Newtonian fluids, the delayed transition from the uniform film to the parabolic region characterized by a constant dimensionless curvature becomes significant due to the higher effective shear rate in the film. Considering the viscosity field obtained from the numerical simulations by \citet{moreira2020isolated}, the appearance of high viscosity in the film is due to the almost stagnant liquid, while in the axis of the channel it is due to a low velocity gradient. In between, as $Cu$ can be interpreted as the ratio between the representative shear rate in the film to the onset of the shear-thinning effects, larger values of $Cu$ (i.e. lower $El$) indicate stronger shear-thinning effects and thus the weight of the second term on the left-hand side of equation (\ref{ODE}) increases, resulting in the delayed transition to the parabolic region for the bubble shape. A similar trend is observed when decreasing $n_c$ at fixed $Cu=13.8\pm0.8$ as shown in Figure \ref{front}(\textit{b}). The decrease of $n_c$ indicates stronger shear-thinning effects and thus the transition to the parabolic region is also expected to be delayed. We note that the numerical solution of equation (\ref{ODE}) starts to deviate from the experimental data as $Ca_0 > 0.3$ (Figure \ref{front}(\textit{b})) since the lubrication approximation will no longer strictly hold for the relatively large $Ca_0$. We note that such a delayed transition from the thin film to the parabolic profile is also theoretically observed in the case where a charged oil droplet moves through a charged capillary. When the electrostatic interaction between the capillary wall and the droplet surface is attractive, the visco-electro-osmotic balance might not only reduce the film thickness, but also delay the transition because of the cooperation of the electro-osmotic and capillary pressure \citep{grassia2020viscous,grassia2022electro}.
 
For the bubble rear meniscus, equation (\ref{ODE}) is solved following the initial and boundary conditions given by \citet{picchi2021motion}. The general observation of the rear meniscus is similar to that of the Newtonian case, where the bubble profile exhibits one main crest and one main valley \citep{magnini2017pore} with a high degree of undulations as shown in Figure \ref{rear}. Although the experimental results do not agree with the numerical solutions very well, the general trends are consistent with the results in \citet{picchi2021motion}, considering the effect of the shear-thinning rheology. As the shear-thinning effect becomes more important with increasing $Cu$ or decreasing $n_c$, the bubble profile stretches along $\xi$. Unlike the front meniscus, where the viscosity profile is regular, the axial velocity gradient near the rear meniscus should be considered when computing the shear rate due to the undulations \citep{picchi2021motion}. Therefore, an accurate description of the viscosity field at the rear meniscus of the bubble requires a further correction for the axial derivative of the velocity in future work.


\section{Conclusion}\label{sec:filetypes}
In this work, we provide an experimental framework to study the motion of a long bubble translating in a circular capillary tube filled with non-Newtonian shear-thinning fluids. The Carreau-Yasuda rheological model is used to describe the rheological properties of the CMCell and Carbopol solutions, with full consideration for the viscosity plateaus at the very low- or high-shear-rates and the shear-thinning behavior at the intermediate-shear-rates. We show that the deposited film thickness and the bubble speed cannot be scaled by the modified capillary number based on the simple power-law rheological model. Instead, the extended Bretherton's law holds well if the effective capillary number is considered, as a function of the Carreau number and power-law index in the Carreau-Yasuda rheological model. In addition, we investigate the shear-thinning effect on the variation of the bubble profile near the front and rear menisci. Based on a recent theoretical work by \citet{picchi2021motion}, we systematically compare the experimental measurements to the numerical prediction for the bubble profile. Stronger shear-thinning effect, indicated by large Carreau number and smaller power-law index, delays the transition from the uniform film to the parabolic region at the bubble front while stretches the undulations at the rear meniscus. The numerical prediction of the bubble profile works well for the bubble front while with less accuracy for the bubble rear given the complexity of the velocity field. We believe our results serve as an experimental validation of the recent modeling approach, which assists the confidence of applying these models to a variety of problems involving lubrication and coating flows with shear-thinning fluids. The influences of other rheological properties, such as viscoelasticity and the resulting coupling with the channel geometry, on the deposition dynamics will be the focus of our future investigation.

\section*{Acknowledgement}
We acknowledge Prof. Randy H. Ewoldt and Yilin Wang in Mechanical Science and Engineering at the University of Illinois at Urbana-Champaign for fruitful discussion about the rheology analysis. Rheological experiments were carried out in part in the Materials Research Laboratory Central Research Facilities, University of Illinois. We also acknowledge the support from American Chemical Society Petroleum Research Fund Grant No. 61574-DNI9 (to J.F.).

\section*{Declaration of interest}
The authors report no conflict of interest.

\bibliographystyle{jfm}
\bibliography{jfm-instructions}

\end{document}